\documentclass[pdflatex,sn-basic,iicol]{sn-jnl}% Basic Springer Nature Reference Style/Chemistry Reference Style

\usepackage{graphicx}%
\usepackage{multirow}%
\usepackage{amsmath,amssymb,amsfonts}%
\usepackage{amsthm}%
\usepackage{mathrsfs}%
\usepackage[title]{appendix}%
\usepackage{xcolor}%
\usepackage{textcomp}%
\usepackage{manyfoot}%
\usepackage{booktabs}%
\usepackage{algorithm}%
\usepackage{algorithmicx}%
\usepackage{algpseudocode}%
\usepackage{listings}%
\usepackage{upgreek}
\usepackage{subfig}
\usepackage{lineno}
\def\path{\string~/}
\usepackage{bm}
\usepackage{pgfplots}
\usepackage{tikz}
\usetikzlibrary{shapes,shadows,arrows,positioning,patterns,decorations.pathreplacing,calc}

\begin{document}

\title{Spatially resolved measurement of the electrostatic charge of turbulent powder flows}

\author[1]{\fnm{Wenchao} \sur{Xu}}\email{wenchao.xu@ptb.de}
\author[1]{\fnm{Simon} \sur{Janta\v{c}}}\email{simon.jantac@ptb.de}
%\author[1,2]{\fnm{Gizem} \sur{Ozler}}\email{gizem.ozler@ptb.de}
\author[2]{\fnm{Tatsushi} \sur{Matsuyama}}\email{tatsushi@t.soka.ac.jp}
\author*[1,3]{\fnm{Holger} \sur{Grosshans}}\email{holger.grosshans@ptb.de}

\affil*[1]{\orgdiv{Analysis and Simulation in Explosion Protection}, \orgname{Physikalisch-Technische Bundesanstalt (PTB)}, \orgaddress{\street{Bundesallee 100}, \city{Braunschweig}, \postcode{38116}, \country{Germany}}}
\affil[2]{\orgdiv{Faculty of Engineering}, \orgname{Soka University}, \orgaddress{\city{Hachioji}, \postcode{192-8577}, \country{Japan}}}
\affil[3]{\orgdiv{Institute of Apparatus and Environmental Technology}, \orgname{Otto von Guericke University of Magdeburg}, \orgaddress{\street{Universit\"{a}tsplatz 2}, \city{Magdeburg}, \postcode{39106}, \country{Germany}}}

\abstract{
This article reports on measurements of the electrostatic charge of particles in a turbulent duct flow.
In contrast to previous charge measurements, which do not apply to turbulent flows or give only the sum of all particles' charges, the new method resolves the charge of a turbulent powder flow spatially.
The experiment consists of a Particle Tracking Velocimetry (PTV) system and electrode plates that generate an electric field.
By comparing particle velocities and accelerations with and without the electric field, the time-averaged local particle charge profile is derived.
Spatially resolving the charge profiles unveiled bipolar particle flow.
The average of the charge profiles agreed well with a conventional Faraday pail measurement, demonstrating the accuracy of our measurements.
However, the peak value of the charge profiles was 76 times higher than the average of the particles' charge.
}


\maketitle

\section{Introduction}

In the UK and Germany, undetected static electricity causes one dust explosion every ten days~\citep{Glor03}. 
Out of all industrial powder operations, pneumatic conveying by turbulent duct and pipe flows leads by far to the highest charge~\citep{Kli18}.
Even though the understanding of powder flow charging in pneumatic conveyors was extensively progressed by Faraday pails~\citep{Nda11,Pel18,Gro23a}, they cannot spatially resolve the charge.
Therefore, local charge peaks of both polarities remained hidden.

Further, simulations of powder flow charging remain immature due to a lack of detailed validation data \citep[see the review of][]{Gro23d}.
Direct numerical simulations nowadays resolve the powder charge profiles in ducts, pipes, and channels~\citep{grosshans2017Direct,Zheng23}.
These spatially highly-resolved simulations revealed the small-scale mechanisms determining the powder flows' charging rate.
For example, particle-bound charge transport led to vigorous charging of highly inertial particles and inter-particle charge diffusion to the charging of low inertial particles.
The validation and improvement of these codes urgently require spatially resolved experiments.

Powder flows charge when particles contact surfaces or other particles, or without contact, for example, by absorbing ions from the atmosphere~\citep{Gou09}.
When a Faraday pail encloses the flow, the charge of the particles inside the pail induces an equal charge on its conductive surface.
The induced charge can be displayed by an electrometer.
The Faraday pail gives the algebraic sum of the enclosed particles' charge, or their average charge if the number of particles is known.

The Faraday pail's advantage is its simplicity and price;
its disadvantage is that it may detect only a minute fraction of the total charge.
For example, through contact with other particles of the same material, powder charges bipolar~\citep{Wait14,Kon17}.
A bipolar powder comprises particles of both polarities, while their overall charge can remain neutral.
In a turbulent flow, mid-sized particles obtain the highest negative and large ones the highest positive charge~\citep{Gro23c}.
If the powder within the Faraday pail's measurement volume is bipolar, the detected charge may be close to zero.

For the same reason, a Faraday pail cannot resolve particle charge distributions, spatial charge profiles, or any other detailed quantity.
Thus, local charge peaks and their possible discharges continue to threaten process safety.

Alternatively, when the conveying pipe is grounded, the charge transferred from the particles to the wall can be detected as electric current~\citep{Mat06,Tagh20}.
Likewise, this signal represents the sum of the charge without any further resolution.

A different flow charge measurement technology applies an external electric field to the powder.
The field separates particles of different polarity~\citep{Toth17} or deflects the particle trajectories~\citep{Maz91,Ham19}.
From the balance of forces acting on the individual particles, the deflected trajectories tell the particles' charge.

The problem of those force balance methods is that they do not apply to turbulent flows.
Solving the force balance requires the knowledge of the flow velocity at the location of the particles, which is readily available in a vacuum, still fluid, or laminar flow.
In turbulence, chaotic fluid forces affect the particles' trajectories.
Thus, the trajectories do not tell the particles' charge or polarity.

The simultaneous measurement of the particles' trajectories and the turbulent flow velocity requires seeding the fluid with tracers.
Seeding tracers to the flow is usually impractical in industrial conveying systems since it interrupts the operation and contaminates the product.
%After collecting a particle sample, these devices can measure the charge distribution, but they are unsuitable for online measurements.

Another force balance approach uses a periodically oscillating electric field.
From the oscillation amplitude of a responding particle, its charge can be deduced.
This technique has been applied to free-falling cloud droplets of a size of 10~$\upmu$m to 100~$\upmu$m~\citep{twomey1956Electrification,takahashi1973Measurement,wells1919oscillation}.
The droplets fall through quiescent air, which enables solving the force balance.
However, the lack of fluid forces prevents applying the method to turbulent flows.
Thus, today's force balance approaches do not apply to turbulent flows where the fluid forces on the particles are unknown.

%Extending the method to turbulent flows would require separating the particle oscillations induced by the electric field from those by turbulent fluctuations.
%In the mentioned studies, the electric field oscillates at a frequencies of 50~Hz or 60~Hz, corresponding to the power line frequencies in most world regions.
%In turbulent duct flows, the fluid fluctuates at frequencies of several magnitutes higher, which prevents isolating the particle oscillations caused by the electric field.
%Further, due to the high streamwise velocity in pneumatic conveying, the particles travel large distances downstream during each oscillation of the electic field, which causes loss of spatial resolution.

Induction probes mounted at towers resolved the internal electrification of dust storms~\citep{Zhang20}, but scaling down the probes to industrial flows would heavily intrude on the hydrodynamics.
Therefore, in fluidized beds, induction probes spatially resolved the charge profile at the boundary walls~\citep{Gaj85}, while the charge within the flow is unknown.
%The signal of induction probes, on the other hand, is difficult to interpret in turbulent flows.

%consequences
A methodology to measure the charge profile of powder conveyed in turbulent duct, pipe, and channel flows is unavailable today.
Consequently, state-of-the-art experimental data is limited to reveal global trends, e.g., the relation between the flow velocity, solid mass loading, or particle material with the total powder charge.
Local charge peaks remain undetected by current measurement technology.
Moreover, bi-polar charging of same-material particles was found in vacuum \citep{Wait14}, but due to the lack of a measurement method, not in turbulent flow.
This lack hinders safety evaluations, particularly of pneumatic conveyors, and the validation of simulations.

%new technology
To measure the powder charge profiles in a turbulent duct flow, we invented and patented an in situ, laser-based measurement technology~\citep{Gro22a}.
This new technology can spatially resolve the time-averaged particle charge profile, including flow regions of opposite polarities.
In this paper, we communicate the realization of the technology in our lab and the first successful measurements.

\section{Experimental method}

\subsection{Test-rig \& measurement section}
\label{sec:rig}

\begin{figure*}[tb]
\centering
\subfloat[]{
\begin{tikzpicture}[scale=0.44]\label{fig:poexp_a}
\draw [->,>=latex,line width=2] (-1,8.5) node [left] {particle feeder} to [out=0,in=90] (0.5,7);
\path [fill=gray] (.6,6) -- (1.2,7) -- (-0.2,7)-- (.4,6);
\path [fill=gray] (.4,6.0) rectangle (.6,5.8);
\path [shading = axis,rectangle, left color=blue!30!white, right color=blue!60!white,shading angle=135] (0,-1) rectangle (1,5.8);
\node [left] at (-1,1) {measurement section};
\path [fill=green,opacity=.4] (-.5,0.8) rectangle (1.5,1.2);
\path [preaction={fill=black!30!red,opacity=0.7},pattern=crosshatch,pattern color=black!70!red,opacity=0.7] (0,.5) rectangle (-.1,1.5);
\path [preaction={fill=black!30!red,opacity=0.7},pattern=crosshatch,pattern color=black!70!red,opacity=0.7] (1,.5) rectangle (1.1,1.5);
\draw [-|,thick] (2,1.5) -- (2,5.8) node[midway,right]{33\,$D$};
\draw [-|,thick] (2,.8) -- (2,1.5) node[midway,right]{2\,$D$};
\draw [|-|,thick] (2,-1) -- (2,.8) node[midway,right]{6\,$D$};
\draw [->,>=latex,thick] (-.7,5) -- (-.7,3.5) node [below] {$x$};
\draw [->,>=latex,thick] (-1,-.25)  node [left]{$D$ = 50 mm} -- (0,-.25);
\draw [<-,>=latex,thick] (1,-.25) -- (1.3,-.25);

\path [rounded corners=2pt,fill=gray!20] (-2,-2) rectangle (1,-3.7) node[align=center,midway]{cyclone\\\& filter};
\draw [rounded corners=2pt,thick] (-2.3,-1) rectangle(1.3,-4) node [midway,yshift=-25] {Faraday pail};
\draw [->,>=latex,line width=2] (.5,-1) to (.5,-2);
\draw [->,>=latex,line width=2] (-2.5,-2.5) to (-4,-2.5) node [left,align=center] {blower};
\end{tikzpicture}
\label{fig:poexpa}
}
\hfill
\subfloat[]{
\raisebox{5mm}{
\begin{tikzpicture}[scale=0.4]
\path [rounded corners=5pt,shading = axis,rectangle, left color=blue!30!white, right color=blue!60!white,shading angle=135] (0,0) rectangle (8,8);
\path [rounded corners=5pt,fill=white] (0.2,0.2) rectangle (7.8,7.8);

\path [preaction={fill=black!30!red,opacity=0.7},pattern=crosshatch,pattern color=black!70!red,opacity=0.7] (0,0) rectangle (-.2,8);
\path [preaction={fill=black!30!red,opacity=0.7},pattern=crosshatch,pattern color=black!70!red,opacity=0.7] (8,0) rectangle (8.2,8);

\draw [-<,>=latex,thick] (0,0) to [out=-20,in=180] (4,-.5); \draw [thick] (3.9,-.5) to [out=0,in=200] (8,0);
\draw [-<,>=latex,thick] (0,8) to [out=20,in=180] (4,8.5); \draw [thick] (3.9,8.5) to [out=0,in=160] (8,8);
\foreach \i in {0.5,...,7.5} \draw [-<,>=latex,thick] (0,\i) -- (4,\i);
\foreach \i in {0.5,...,7.5} \draw [,thick] (0,\i) -- (8,\i);

\draw [->,>=latex,ultra thick] (-1,4.4) -- (-1,5.5); \draw [->,>=latex,ultra thick] (-1,3.6) -- (-1,2.5);

\draw [green,line width=6,opacity=.4] (-1.5,4) -- (9.5,4);
\path [fill=blue] (3.5,11) rectangle (4.5,12.5) node [midway,above,yshift=8] {camera};
\path [fill=blue] (3.9,10.8) rectangle (4.1,11);
\node at (6,2) {$E_y$};
\draw [<->,>=latex,thick] (0,9.5) node [right] {$y$} -- (-1.5,9.5) -- (-1.5,11) node [above] {$z$};
\end{tikzpicture}
\label{fig:poexpb}
}}
\hfill
\subfloat[]{
\raisebox{5mm}{
\begin{tikzpicture}[scale=0.4]
\path [shading = axis,rectangle, left color=blue!10!white, right color=blue!40!white,shading angle=135] (0,-1.5) rectangle (8,7.5);
\draw [<->,>=latex,thick] (0,10) node [right] {$y$} -- (-1.5,10) -- (-1.5,8.5) node [below] {$x$};
\path [preaction={fill=black!30!red,opacity=0.7},pattern=crosshatch,pattern color=black!70!red,opacity=0.7] (0,0) rectangle (-.2,6);
\path [preaction={fill=black!30!red,opacity=0.7},pattern=crosshatch,pattern color=black!70!red,opacity=0.7] (8,0) rectangle (8.2,6);
\path [fill=green,opacity=0.4] (-1.5,1) rectangle (9.5,5);
\draw [gray,-<,>=latex,thick] (0,6) to [out=20,in=180] (4,6.5); \draw [gray,thick] (3.9,6.5) to [out=0,in=160] (8,6);
\draw [gray,-<,>=latex,thick] (0,0) to [out=-20,in=180] (4,-.5); \draw [gray,thick] (3.9,-.5) to [out=0,in=200] (8,0);
\foreach \i in {0.5,...,5.5} \draw [gray,-<,>=latex,thick] (0,\i) -- (4,\i);
\foreach \i in {0.5,...,5.5} \draw [gray,thick] (0,\i) -- (8,\i);
\coordinate (a) at (3.2,4.4);
\coordinate (b) at (3,3);
\coordinate (c) at (2.8,1.6);
\coordinate (d) at (3.7,3);
\coordinate (e) at (4.5,1.6);
\draw [->,>=latex,thick,shorten >=1mm] (a) -- (b);
\draw [->,>=latex,thick,shorten >=1mm] (b) -- (c);
\draw [->,>=latex,thick,shorten >=1mm,red] (a) -- (d);
\draw [->,>=latex,thick,shorten >=1mm,red] (d) -- (e);
\shade [ball color=black] (a) circle (2mm) node [above] {${\bm x}(t_1)$};
\shade [ball color=black] (b) circle (2mm) node [left] {${\bm x}(t_2)$};
\shade [ball color=black] (c) circle (2mm) node [below left] {${\bm x}(t_3)$};
\shade [ball color=red] (d) circle (2mm) node [right,red] {${\bm x}^E(t_2)$};
\shade [ball color=red] (e) circle (2mm) node [below right,red] {${\bm x}^E(t_3)$};
\end{tikzpicture}
\label{fig:poexpc}
}}
\caption[]{
(a) Pneumatic conveying pilot plant to test the new measurement system.
(b) Components of the measurement system: electric field ($E_y$), traversable laser sheet (green), and PTV camera.
(c) Motion of a particle from ${\bm x}(t_1)$ with the $E$-field to ${\bm x}^E(t_3)$ and without the $E$-field to ${\bm x}(t_3)$.
}
\label{fig:poexp}
\end{figure*}
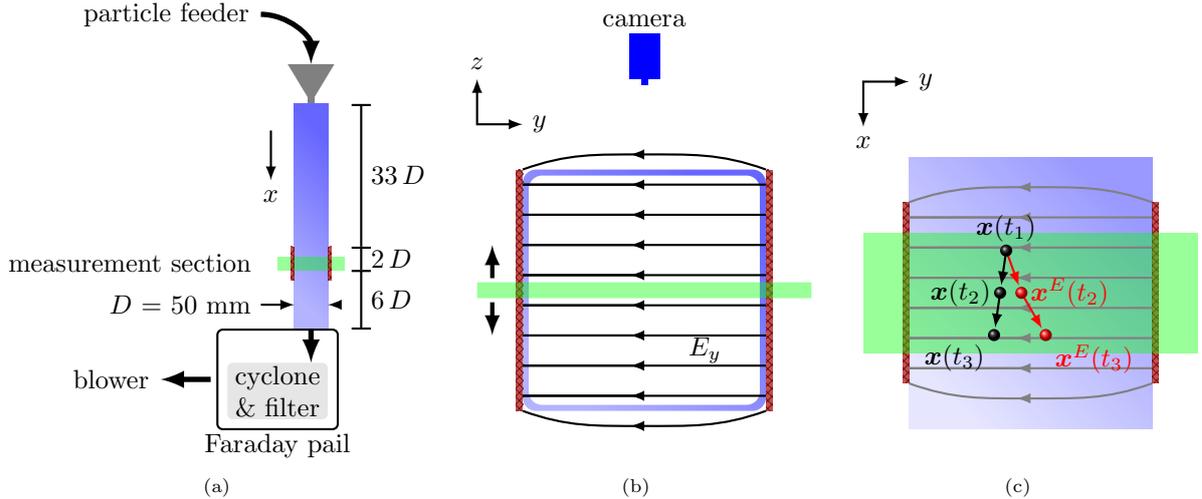

For developing the measurement technology, we installed a pneumatic conveying pilot plant (figure~\ref{fig:poexp_a}) in our lab.
A mini screw feeder manufactured by SEIWA GIKEN supplies particles to the duct's top.
The duct conveys the powder vertically downward, aligned with the gravitational acceleration.
After the duct's outlet, a cyclone and a filter separate the particles from the airflow.
A blower sucks the air and powder through the duct, cyclone, and filter.
Between the blower and the filter, an orifice airflow meter O-METER type OM4S measures the airflow rate.

A conventional Faraday pail provides validation data for the new measurement method.
The pail encloses the cyclone and the filter, while their metallic housings are connected to a Keithley~6514 electrometer.
Thus, the electrometer measures the sum of the charge of the particles collected by the cyclone and the filter.

The duct is made of transparent PMMA (polymethyl methacrylate), providing optical access for Particle Tracking Velocimetry (PTV) (figures~\ref{fig:poexpb} and \ref{fig:poexpc}).
To avoid the laser beam's diffraction at curved surfaces, the duct is of a square cross-section.
Each planar wall's inner side length is $D=$\,50\,mm, and its thickness is 0.16~$D$.
The duct's length, from the inlet to the beginning of the measurement section, is 33~$D$.
The flow in the duct is statistically stationary.

\begin{figure}[tb]
\centering
\includegraphics[trim=0mm 0mm 0mm 0mm,clip=true,width=0.47\textwidth]{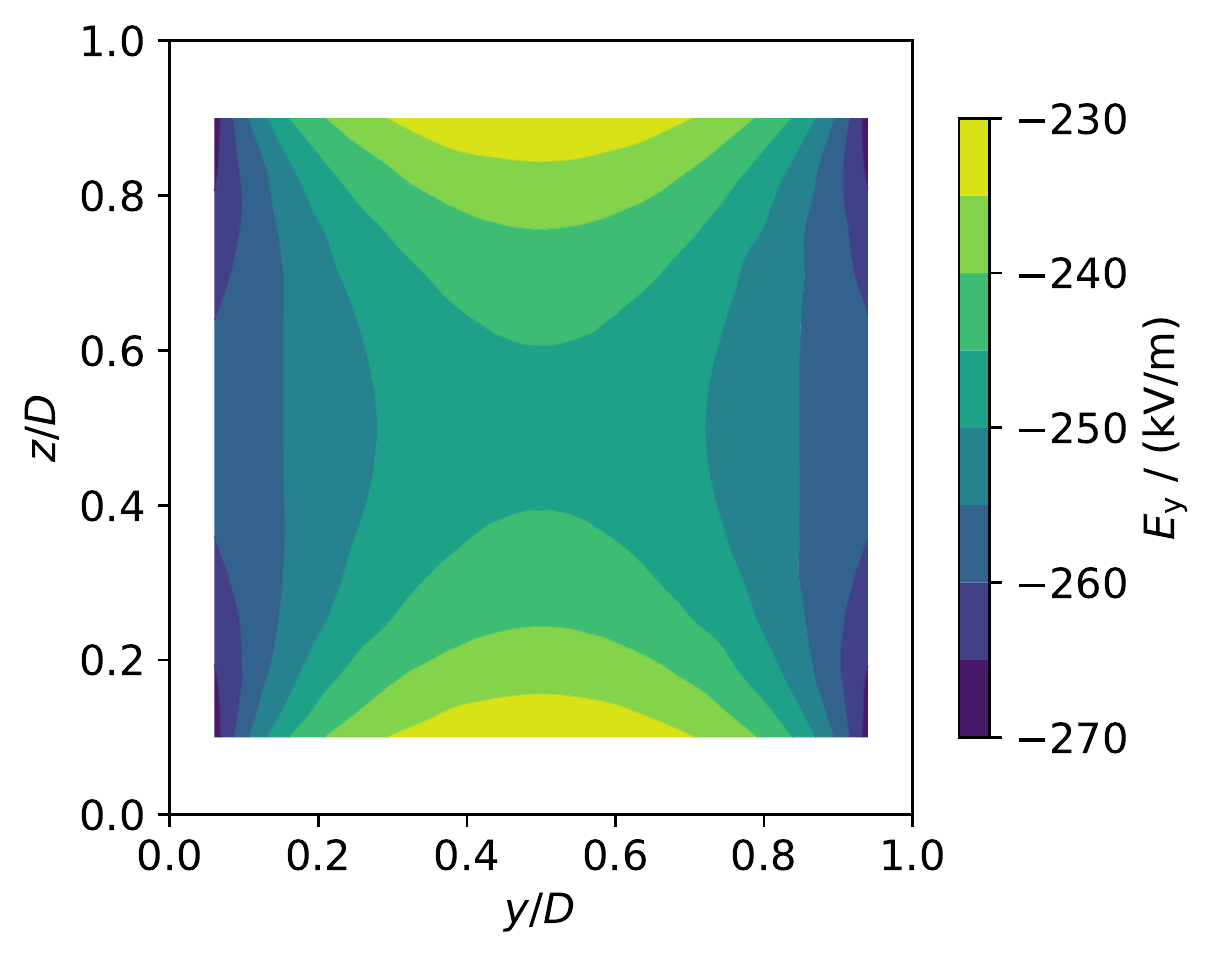}
\caption[]{
Distortion of the electric field in $y$ direction by the PMMA duct, computed with COMSOL.
}
\label{fig:ef}
\end{figure}

In the measurement section, at opposite duct walls, two parallel electrodes connected to a high-voltage power supply generate an electric field.
To avoid contact with the particles, the electrodes are attached to the outside of the duct.
They are manufactured of thin, transparent Indium-Tin-Oxide (ITO) coated plastic films that let the laser beam pass through.
The negatively charged electrode is located at $y/D=-0.16$ and the positive one at $y/D=1.16$. 
The enclosed PMMA duct distorts the electric field.
Figure~\ref{fig:ef} shows the $y$ component of the electric field computed with COMSOL.

The electric field can be switched on or off.
When the electric field is switched on, it accelerates positively charged particles in the $-y$ and negatively charged particles in the $+y$~direction. 
The measurement system analyses the particles' response to the electric field.

A 40\,W pulse diode (LD-PS) laser illuminates the flow at a wavelength of 450\,nm\,$\pm$\,20\,nm.
The laser emits repetitive pulses, each lasting 34.5~$\upmu$s, at a frequency of 4353 s$^{-1}$.

The laser beam points in $y$~direction.
Lenses shape the beam into a sheet that spreads across the $x$-$y$ plane.
In the $z$ direction, the laser's intensity profile has an effective thickness of about 2~mm along the whole camera field of view.
In the streamwise direction ($x$), the sheet begins $D$ after the edge of the electric field spans $D$ downstream.
To scan the complete cross-section of the duct, the laser sheet is traversable in the $z$~direction (see figure~\ref{fig:poexpb}).

The gas flow's bulk Reynolds number was 13\,200, based on $D$ and the average velocity.
The particles were monodisperse, spherical, of a size of $d=100$\,$\upmu$m, and made of PMMA.
According to the manufacturer, the standard deviation of the particle size was less than 3.5\,$\upmu$m.
The powder mass flow rate was 0.2\,g/s\,$\pm$\,0.02\,g/s, corresponding to an average particle number density of about $3.3\times 10^7$\,m$^{-3}$, solid volume fraction of $1.7 \times 10^{-5}$, or mass fraction of $1.7 \times 10^{-2}$.
The relative humidity and temperature in the lab during the tests were 47\% and 19.5~$^\circ$C.

\subsection{PTV setup}

A 2D2C particle tracking velocimetry (PTV) system captured the particles' motion within the measurement section.
Part of the PTV system is a Chronos~2.1 monochrome high-speed camera operating at a frame rate of 4352\,fps.
At the applied frame rate, the camera has a resolution of 800~$\times$~600 pixels, each pixel of a size of 100~$\upmu$m~$\times$~100~$\upmu$m.
At this resolution and frame rate, the camera's internal memory allows a maximum recording duration of 10.94\,s, yielding a total of 47\,610 frames.
Thus, the data presented in this paper were averaged over 47\,609 time intervals.

For initial image processing, a filter based on proper orthogonal decomposition (POD) eliminated static background elements \citep{mendez2017PODbased}.
This filter removes bright artifacts, such as the duct's illuminated sidewall and reflections of contaminations on the inner wall, thus mitigating errors in the near-wall area of the images.

To extract particle trajectories, the PTV system identifies and links image features using the open-source Python package Trackpy \citep{allan_2023}.
Trackpy is based on the tracking algorithm by \citet{crocker1996Methods}.
Further, it achieves sub-pixel precision through the least-squares fitting method \citep{vanderwel2017Automated}.

Since the method described below to derive the particle charge requires precise acceleration data, we stringently filtered spurious trajectories.
The filter retains only particle trajectories identified on at least 7 subsequent frames and discards any trajectory on fewer frames.
Those trajectories that met this criterion were further refined by a B-spline smoothing filter \citep{eilers1996Flexible} that removes irregularities from the data.
Also, the filter rejected the trajectories for which the residuals of the smoothing were above a threshold (see section~\ref{sec:acc}).

\subsection{Deriving $Q$ from the force balance}
\label{sec:math}

The force balance of a charged particle in an electric field in $y$ direction, the longitudinal axis of the laser beam, reads
\begin{equation}
\label{eq:ae}
m a^E_y = F_{\mathrm{d},y} + F_{\mathrm{c},y} + F_{\mathrm{e},y} \, , 
\end{equation}
where $m$ and $a^E_y$ are the particle's mass and wall-normal acceleration.
The right-hand-side of the above equation sums the forces acting on the particle, the drag force, $F_{\mathrm{d},y}$, collisional force, $F_{\mathrm{c},y}$, and electrostatic force, $F_{\mathrm{e},y}$.
The collisional force term includes the forces on a particle during contact with a duct wall or other particles.
Gravitational forces vanish since the laser points horizontally.

The drag term in $y$ direction is
\begin{equation}
\label{eq:fd}
F_{\mathrm{d},y} = m \dfrac{u_{\mathrm{g},y} - u^E_y}{\tau} \, ,
\end{equation}
where $u_{\mathrm{g},y}$ is the free-stream velocity of the gas surrounding the particle and $u^E_y$ the particle's velocity.
%The particle response time, $\tau$, quantifies the time the particle takes to adapt to a change of the gas velocity.
Assuming a Stokes flow, the response time, 
\begin{equation}
\label{eq:tau}
\tau= \rho \, d^2/(18 \, \mu) \, ,
\end{equation}
is constant.
Herein, $\rho=$~1150~kg\,m$^{-3}$ is the particles' material density and $\mu=$~1.825\,$\times$\,10$^{-5}$\,N\,m$^{-2}$ the dynamic gas viscosity.
The Stokes flow assumption is valid for a particle Reynolds number of $Re_\mathrm{p}=u_\mathrm{rel}\,d/\nu < 1$, where $u_\mathrm{rel}$ the magnitude of the relative velocity of the particle and the surrounding air, and $\nu=$~1.516\,$\times$\,10$^{-5}$\,m$^2$\,s the kinematic gas viscosity.

The electrostatic force,
\begin{equation}
\label{eq:fe}
F_{\mathrm{e},y} = Q \left( E_y + E_{\mathrm{oth},y} \right) \, ,
\end{equation}
depends on the particle's charge, $Q$, and the electric field's component in $y$ direction.
The total electric field in the measurement section comprises $E_y$, generated by the electrodes, and $E_{\mathrm{oth},y}$, the sum of other electric fields, for example, by surrounding charged particles, charge located at the duct walls, or mirror charges.

%Given the magnitude of the external electric field, which is on the order of $10^5$~V/m and significantly higher than the electric field generated by surrounding charged particles, charge located at the duct walls, and mirror charges (which are on the order of $10^2 - 10^3$~V/m), we can reasonably approximate $E_{\mathrm{tot},y}$ as being solely determined by the external electric field.
%Hence, $E_{\mathrm{tot},y} \approx E_y$.

Substituting equations~(\ref{eq:fd}) and~(\ref{eq:fe}) in~(\ref{eq:ae}), and dividing by the particle's mass yields
\begin{equation}
\label{eq:q}
a^E_y = \dfrac{u_{\mathrm{g},y} - u^E_y}{\tau} + f_{\mathrm{c},y} + \dfrac{Q \left( E_y + E_{\mathrm{oth},y} \right)}{m} \, .
\end{equation}
Out of the terms of this equation, $E_y$ is known, $m_\mathrm{p}$ is known since the particles are monodisperse, $\tau$ is approximated by the Stokes assumption, and $a^E_y$ and $u^E_y$ are measured by PTV.

However, the problem of equation~(\ref{eq:q}) lies in the unknown terms $f_{\mathrm{c},y}$, $E_{\mathrm{oth},y}$, and $u_{\mathrm{g},y}$.
The instantaneous gas velocity is relevant in turbulent flows with particles of \mbox{$\tau\ll\infty$};
but measuring it simultaneously to the particle velocities requires an additional measurement system and seeding tracer particles, which is impossible in most industrial flows.
The collisional force is unknown.
Even though it is generally small in dilute flows, it can be locally large.
In the following, we propose a solution to this problem based on averaging.

Time averaging equation~(\ref{eq:q}) yields
\begin{equation}
\label{eq:q21b}
\bar{a}^E_y = 
\dfrac{\bar{u}_{\mathrm{g},y} - \bar{u}^E_y}{\tau} 
+ \bar{f}_{\mathrm{c},y} 
+ \dfrac{ \bar{Q} E_y + \overline{Q E}_{\mathrm{oth},y} }{m} \, .
\end{equation}
The operator $\bar{\phi}$ denotes the time average over many particles or the electric field at fixed points in space.
In other words, $\bar{Q}$ is the spatially-resolved profile of the arithmetic mean of the local charge distribution,
\begin{equation}
\label{eq:qbar}
\bar{Q} = \bar{Q}(\bm{x}) = \dfrac{1}{N(\bm{x})} \sum^{N(\bm{x})}_{n=1} Q_n(\bm{x}) \, ,
\end{equation}
where $N$ is the number of particles holding a charge $Q_n$ and passing the location $\bm{x}$ during the measurement.

In an experiment with de-activated electric field, equation~(\ref{eq:q21b}) reduces to
\begin{equation}
\label{eq:q22}
\bar{a}_y = 
\dfrac{\bar{u}_{\mathrm{g},y} - \bar{u}_y}{\tau} 
+ \bar{f}_{\mathrm{c},y} 
+ \dfrac{\overline{QE}_{\mathrm{oth},y}}{m} \, ,
\end{equation}
where $\bar{u}_y$ and $\bar{a}_y$ are the particles' velocity and acceleration without the electric field.
If $\bar{u}^E_y$ and $\bar{a}^E_y$ are measured shortly after the particles reach the electric field, the field only slightly affects the particle positions.
Then, in equations~(\ref{eq:q21b}) and~(\ref{eq:q22}) the terms $\bar{f}_{\mathrm{c},y}$ and $\overline{QE}_{\mathrm{oth},y}$ are equal.
Also, because the flow is statistically stationary, the terms $\bar{u}_{\mathrm{g},y}$ are equal.
Thus, substracting equation~(\ref{eq:q22}) from equation~(\ref{eq:q21b}) and solving for $\bar{Q}$ results in
\begin{equation}
\label{eq:q2b}
\bar{Q} = \left( 
\bar{a}^E_y - \bar{a}_y 
+ \dfrac{\bar{u}^E_y - \bar{u}_y}{\tau} 
\right) \dfrac{m}{E_y} \, .
\end{equation}
In other words, in the above equation, the unknown collisional force, electric field components, and gas phase velocity are eliminated even though being implicitly included, which is the advantage of equation~(\ref{eq:q2b}) over equation~(\ref{eq:q}).
In return, equation~(\ref{eq:q2b}) provides the time-averaged charge instead of the charge per particle.

\section{Accuracy of the method}
\label{sec:acc}

\subsection{PTV algorithm and equipment}

In total, 1.8~million recorded particle trajectories passed the PTV algorithm's smoothing filter.
The filter's small threshold keeps the quality of the remaining trajectories high;
the standard deviation of their residual is $1.9\times10^{-3}$ pixels.

The location of the recorded particles distributes over the duct's entire cross-section.
At each location where equation~(\ref{eq:q2b}) was evaluated, $u^E_y$, $u_y$, $a^E_y$, and $a_y$ were averaged over 700 to 5500 measured values. % x1 in plot_eif.py 
The standard deviation of these values propagates to a Standard Error of the Mean (SEM) of the particle charge between 1.4~fC and 3.2~fC, depending on the location, and an average SEM of 2.0~fC. % x2 in plot_eif.py

The PTV camera detects the particle locations with an uncertainty of approximately $1.6\times10^{-3}$ pixels \citep{savin2005Static}, less than 0.2\% of the average particle displacement between subsequent recorded images.
At all locations of the duct's cross-section, this detection error leads to an extremely low SEM of the particle charge ($\ll$~0.1~fC). % x3 in plot_eif.py

According to the manufacturer, the maximum voltage error of the electric field's power supply is 0.5\%.
This error propagates linearly to a maximum error of the electric field and particle charge according to equation~(\ref{eq:q2b}) of 0.5\%. % x4 in plot_eif.py

Further, the uncertainty of the particle diameter (3.5\,$\upmu$m) affects equation~(\ref{eq:q2b}) through the particle mass by the diameter's third-order non-central moment
Thus, the average particle mass is 0.37\% higher than the nominal mass that enters the equation, which leads to a negative offset of 0.37\% of the derived $\bar{Q}$. % x1 in sem.py

\subsection{Stokes flow assumption}

The mathematical derivation of equation~(\ref{eq:q2b}) requires the particle response time to be known and constant.
We satisfied this requirement by assuming a Stokes flow~(equation~(\ref{eq:tau})), which is valid for $Re_\mathrm{p} < 1$.
However, the particles' Reynolds number based on their terminal velocity in still air is approximately 1.5 \citep{Clift78} and exceeds this value instantaneously in turbulence.
Thus, the Stokes flow assumption introduces an error to equation~(\ref{eq:q2b}).

To evaluate the error of the charge of powder in turbulence related to the Stokes flow assumption, we theoretically analyze a single, charged particle moving in $x$ direction ($u_x \gg u_y$ and $u_x \gg u^E_y$) through a uniform and constant flow.
Analogous to the averaged equation~(\ref{eq:q2b}), the charge of a single particle is, based on Stokes' assumption,
\begin{equation}
\label{eq:qSt}
Q_\mathrm{Stokes} = \left(a^E_y - a_y + \dfrac{u^E_y - u_y}{\tau} \right) \dfrac{m}{E_y} \, .
\end{equation}

According to Oseen \citep{Batch10}, who included first-order inertial effects, the particle's response time for Reynolds numbers up to 5 decreases to 
\begin{equation}
\label{eq:oseen}
\tau' = \tau \left( 1+\frac{3}{16} Re_\mathrm{p} \right)^{-1} \, .
\end{equation}
Since the particle moves mainly in the $x$ direction, $Re_\mathrm{p}$ and $\tau'$ are not affected by the change of the particle's velocity in the $y$ direction.
Thus, based on Oseen's correction, the particle's charge is
\begin{equation}
\label{eq:qOs}
Q_\mathrm{Oseen} = \left(a^E_y - a_y + \dfrac{u^E_y - u_y}{\tau'} \right) \dfrac{m}{E_y} \, .
\end{equation}

The error of the particle's charge due to Stokes assumption, $err_Q=Q_\mathrm{Stokes}-Q_\mathrm{Oseen}$, is
\begin{equation}
\label{eq:err1}
err_Q = \left( \dfrac{u^E_y - u_y}{\tau} - \dfrac{u^E_y - u_y}{\tau'} \right) \dfrac{m}{E_y} \, .
\end{equation}
Inserting equation~(\ref{eq:oseen}) into~(\ref{eq:err1}) and reordering the terms yields 
\begin{equation}
\label{eq:err2}
err_Q =
\frac{3}{16}
\dfrac{m}{\tau\,E_y}
\,
Re_\mathrm{p} \left( u_y - u^E_y \right)
\, .
\end{equation}
The last term can be estimated by the velocity in $y$ direction a particle reaches in the electric field, $\left( u_y - u^E_y \right) = \delta t^EQE_y/m$, where $\delta t^E = 2D/u_x$ is the particle's residence time in the measurement section.
Using this estimation and dividing equation~(\ref{eq:err2}) by $Q$ gives the relative error
\begin{equation}
\label{eq:err3}
err_{\mathrm{rel},Q} =
\frac{3}{16}
\dfrac{\delta t^E}{\tau}
Re_\mathrm{p}
\, .
\end{equation}
This expression reveals that for a given $\tau$, i.e., particle diameter, material density, and gas viscosity, the error due to the Stokes flow assumption scales linearly with two factors:
the particles' Reynolds number and residence time in the measurement section.
While $Re_\mathrm{p}$ is a characteristic of the flow, $\delta t^E$ can be reduced by shortening the measurement section.

\begin{figure}[tb]
\centering
\includegraphics[trim=0mm 0mm 0mm 0mm,clip=true,width=0.47\textwidth]{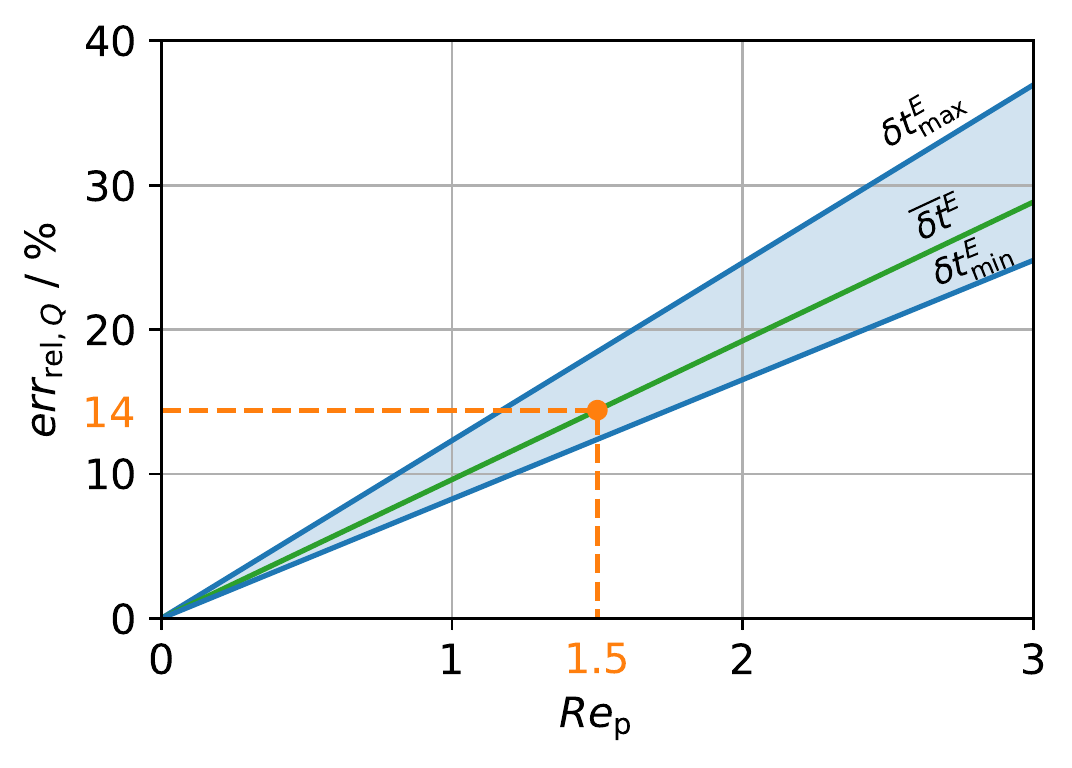}
\caption[]{
Relative error of equation~(\ref{eq:q2b}) stemming from the Stokes flow assumption.
The green line is for the mean residence time in the measurement section, $\overline{\delta t}^E=18$~ms.
The orange symbol marks $Re_\mathrm{p}=1.5$ based on the particles' settling velocity in still air.
}
\label{fig:err_tau}
\end{figure}

Figure~\ref{fig:err_tau} plots equation~(\ref{eq:err3}).
The green line represents the mean $\overline{\delta t}^E=18$~ms determined from the mean downstream velocity, $\bar{u}_x$, of all particles in our experiment. % x1 in .py
The orange symbol marks the error of 14\% based on the particles' settling velocity in still air.
Finally, the lines for $\delta t^E_\mathrm{min}=15$~ms and $\delta t^E_\mathrm{max}=23$~ms depict the error of the particle with the longest and shortest residence time, i.e., the lowest and highest measured $u_x$. % x0 in .py

\subsection{Flow and measurement scales}

% time-scale

Our experimental procedure requires two measurements while the electric field is switched off and on.
To obtain representative average particle velocities and accelerations, the measurements' durations are approximately 840 times the flow time scale $D/\bar{u}_x$.

At the same time, to temporally resolve changes in the mean flow, the complete measurement procedure needs to be short compared to the time scale of boundary condition changes.
Since the flow in the test rig is statistically stationary, this condition is fulfilled.

% spatial scale

Contrary to averaging in time, the measurements resolve the mean flow changes in space.
Thus, there is no limit to the spatial resolution of the method in spanwise ($y$ and $z$) direction.

In the streamwise direction, the length of the measurement section is a compromise between spatial accuracy and a low signal-to-noise ratio.
The signal, the measured velocity and acceleration responses, is the stronger the longer the electric field is.

\begin{figure}[tb]
\centering
\includegraphics[trim=0mm 0mm 0mm 0mm,clip=true,width=0.47\textwidth]{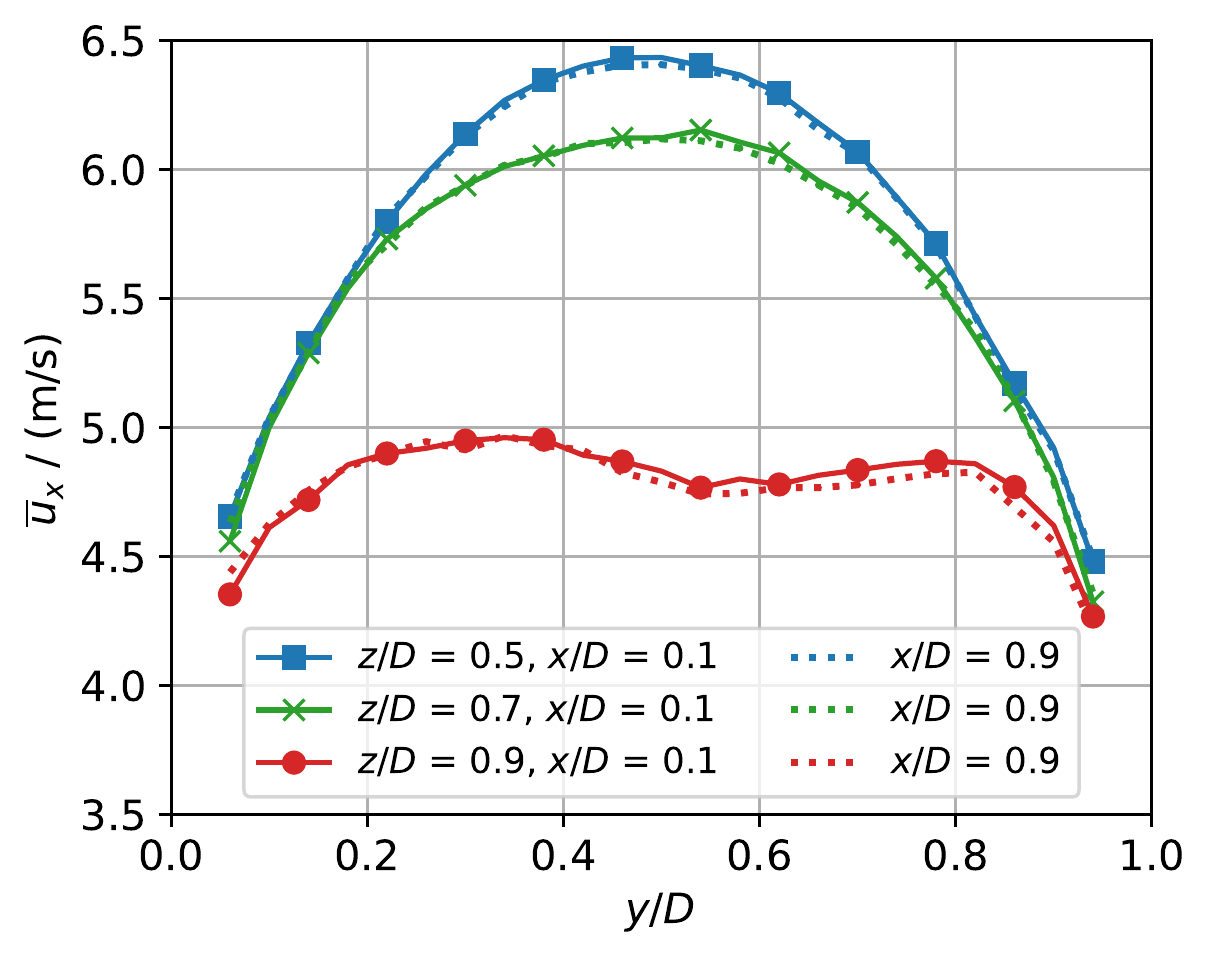}
\caption[]{
Time-averaged streamwise particle velocities.
Comparison of the profiles at the beginning and end of the laser beam for three different slices averaged over $x/D\pm0.1$.
}
\label{fig:ux}
\end{figure}

On the other hand, the measurement section needs to be shorter than the length scale of the changes to the mean flow in the downstream direction.
In our test rig, the flow is not yet fully developed when reaching the measurement section.
However, figure~\ref{fig:ux} shows that the particles' streamwise velocity profiles do not change from the beginning to the end of the laser beam.
Thus, the measurement section is sufficiently short to spatially resolve the flow in the $x$ direction. 

\begin{figure*}[tb]
\includegraphics[trim=0mm 0mm 0mm 0mm,clip=true,width=0.99\textwidth]{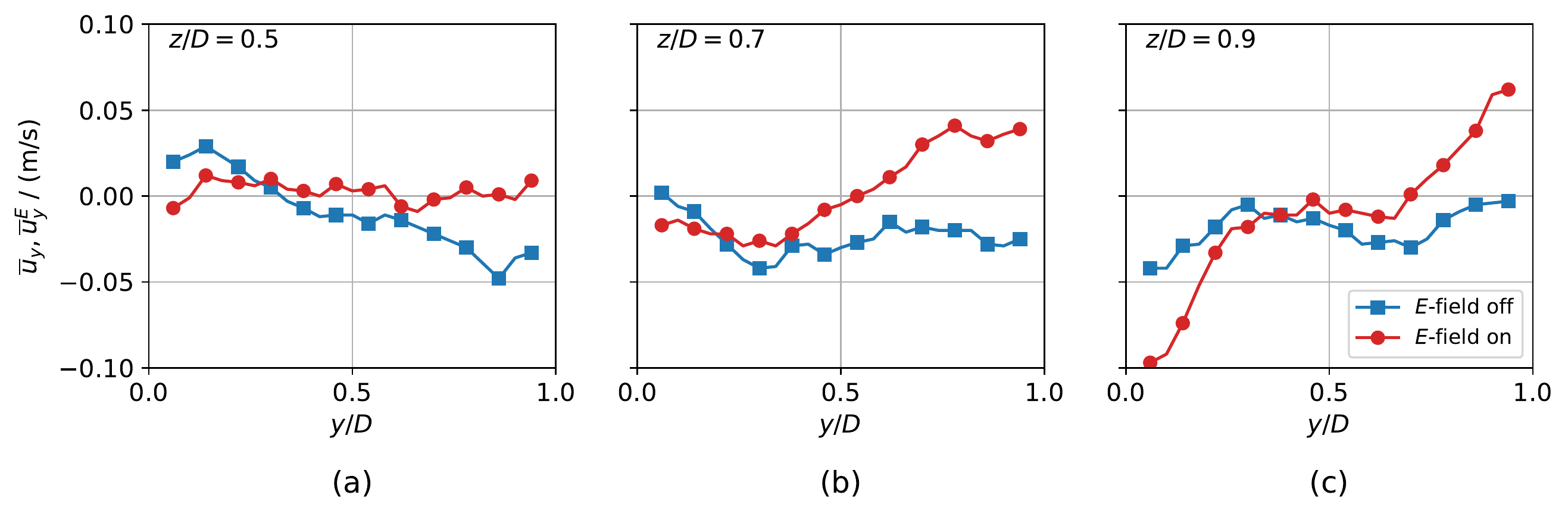}
\caption[]{
Time-averaged wall-normal particle velocities responding to the electric field.
}
\label{fig:u}
\includegraphics[trim=0mm 0mm 0mm 0mm,clip=true,width=0.99\textwidth]{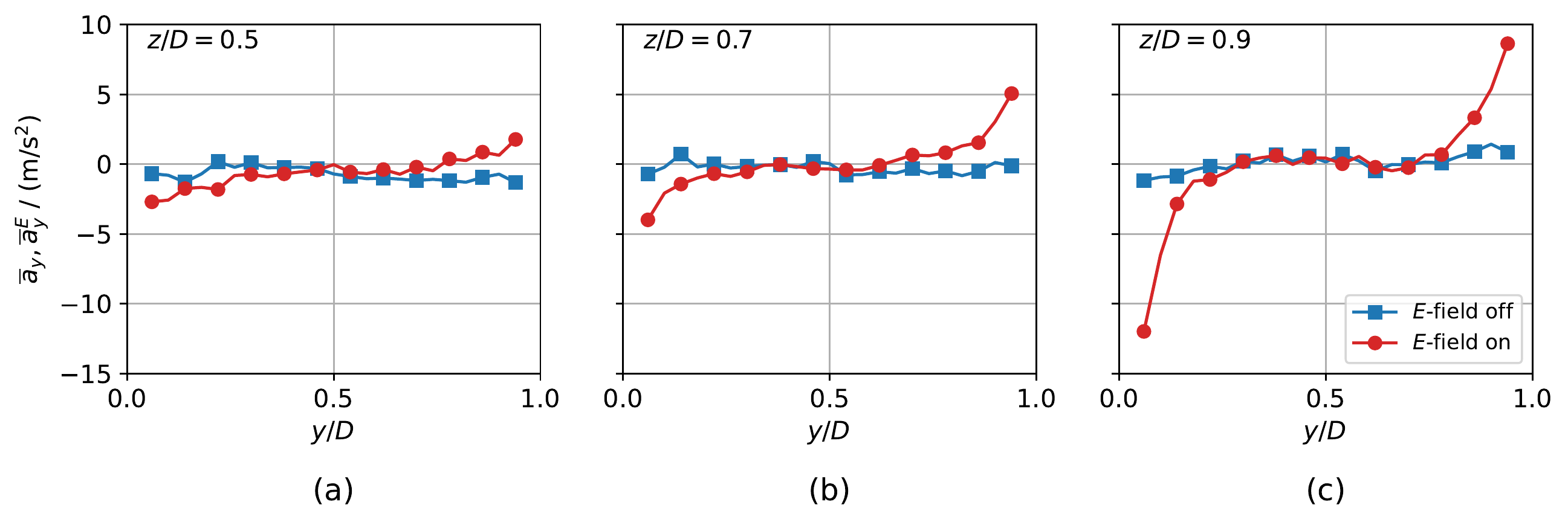}
\caption[]{
Time-averaged wall-normal particle accelerations responding to the electric field.
}
\label{fig:a}
\end{figure*}

% errors: u_e, sorting, f_col

Nevertheless, the electric field gives rise to other errors.
The stronger the particles react to the electric field, the more invasive the measurement.
Analogous to the above estimation of the particles' mean spanwise velocity response, we estimate their mean change of location in $y$ direction due to the electric field to $0.5 (\overline{\delta t}^E)^2 Q E_y /m\approx 0.01\,D$ (assuming $Q=10$~fC). % x2 in errTau.py
Thus, the measurement was slightly invasive.
Moreover, this location change propagates to an uncertainty of the charge profiles' spatial coordinates.

Further, the electric field changes the wall collision frequency.
Thus, close to the walls, $\bar{f}_{\mathrm{c},y}$ in equations~(\ref{eq:q21b}) and~(\ref{eq:q22}) differ, which violates the assumptions underlying equations~(\ref{eq:q2b}).
The product of the particles' residence time and measured spanwise velocity implies that they move less than $|y|<0.03\,D$ within the measurement section. % x3 in errTau.py
Therefore, in the following, we plot data for $0.04<y/D<0.96$, where the results are unaffected by wall collisions.

To sum up, our charge measurements are mainly affected by two errors:
The SEM of the particles' velocities and accelerations and the error of the particles' response time.
Reducing the downstream length of the measurement section can reduce several errors related to spatial uncertainty.
However, the chosen distance ensures a low signal-to-noise ratio of the measured velocity and acceleration responses, which we prioritized over spatial accuracy.

\section{Spatially resolved particle charge}

\begin{figure*}[tb]
%\centering
\includegraphics[trim=0mm 0mm 0mm 0mm,clip=true,width=0.99\textwidth]{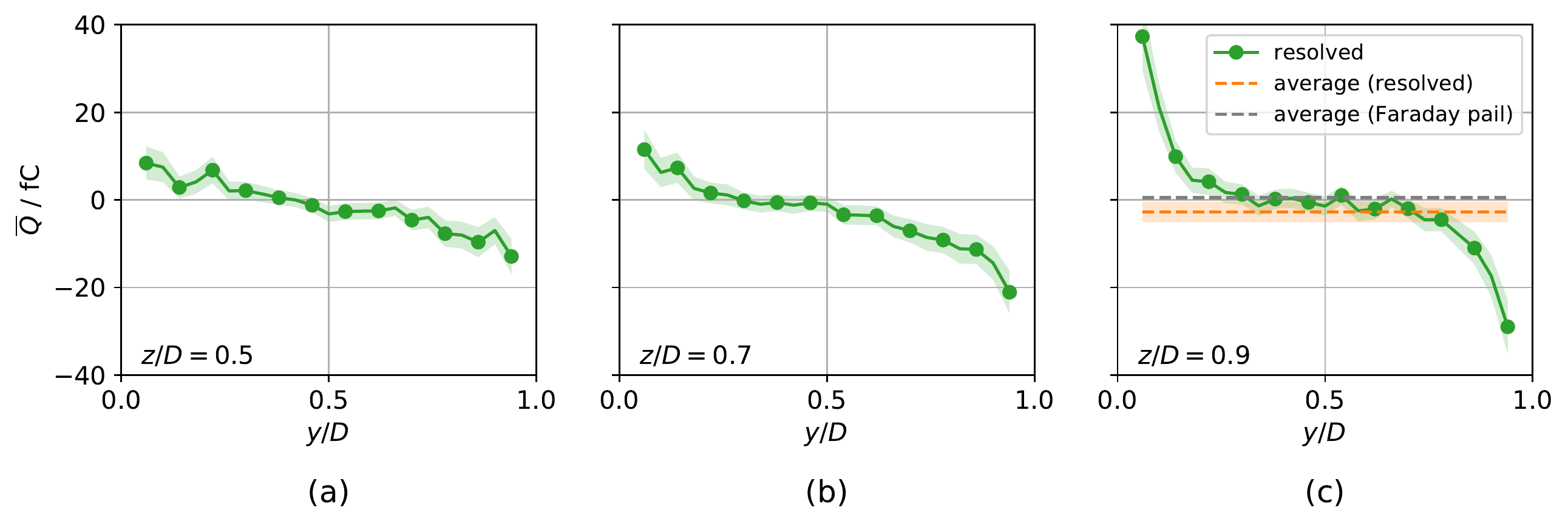}
\caption[]{
Spatially-resolved, time-averaged particle charge over three slices.
The average particle charge of the spatially resolved and the Faraday pail measurements agree well.
}
\label{fig:q_slice}
\end{figure*}

\begin{figure}[tb]
\includegraphics[trim=0mm 0mm 0mm 0mm,clip=true,width=0.50\textwidth]{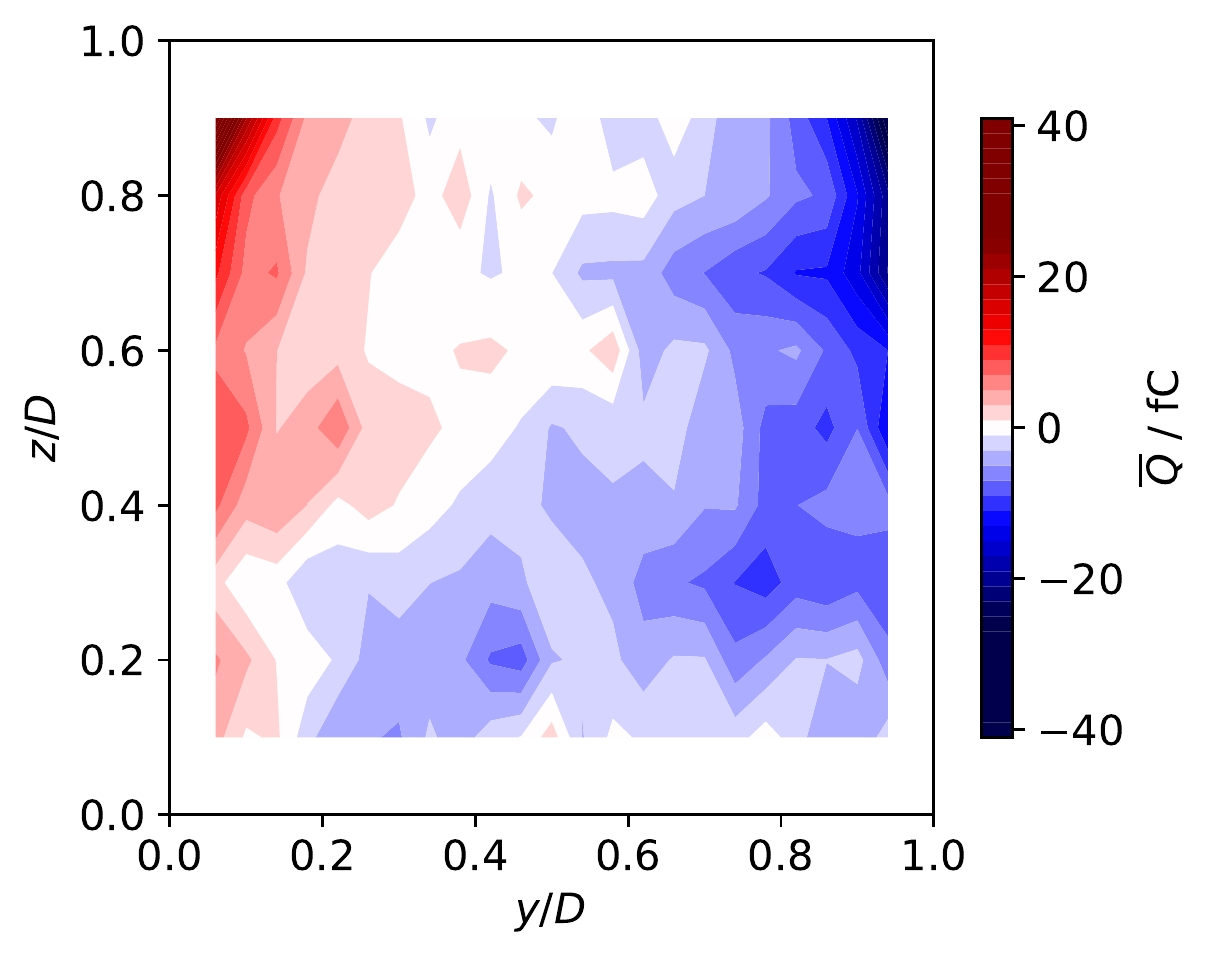}
\caption[]{Spatially-resolved, time-averaged particle charge over the cross-section.}
\label{fig:q_surf}
\label{fig:q}
\end{figure}

Figures~\ref{fig:u} and \ref{fig:a} present the particles' measured wall-normal velocities and accelerations.
All data are time-averaged and depict three spatial profiles in the direction of the electric field.
The velocities and accelerations fill the terms on the right-hand side of equation~(\ref{eq:q2b}) to derive the average particle charge presented in figure~\ref{fig:q}.

Downstream of the point-like feeding position at the duct's center, the particle flow widened toward the walls, as reproduced by the nearly symmetric velocity profile in figure~\ref{fig:u}(a) of the particles without the electric field.
When reaching the measurement section, the flow was not yet fully developed (our measurement technique applies to non-developed flows).
%Also, electric fields ($\sum E_{\mathrm{oth},y}$), even without $E_y$ activated, for example, due to charge spots on the duct, affect the particles.

In all slices, the particle velocity and acceleration respond to the electric field.
The more their profiles differ between the electric field being switched on and off, the more the particles are charged, 

In figures~\ref{fig:u} and \ref{fig:a}, the profiles of the slices closest to the wall ($z/D=$~0.9) respond the strongest to the electric field, and the profile of the slice through the duct's centreline ($z/D=$~0.5) responds the least.
Also, in each slice, the particles close to the walls at $y/D=$~0.0 and $y/D=$~1.0 respond the strongest.

%profiles
Figure~\ref{fig:q_slice} shows the particle charge over three slices of the duct, and figure~\ref{fig:q_surf} over the duct's cross-section.
The shaded areas in figure~\ref{fig:q_slice} give the two main uncertainties affecting the measurements, resulting from the SEM of the particles' velocities and accelerations and the error of the particles' response time (cf.~section~\ref{sec:acc}).
We resolved the charge over the entire cross-section by traversing the measurement plane (cf.~figure~\ref{fig:poexpb}) from $z/D=$~0.1 to $z/D=$~0.9 at intervals of $z/D=$~0.1, while maintaining the flow conditions, and interpolated between the planes.

It is reiterated that figures~\ref{fig:q_slice} and~\ref{fig:q_surf} depict the time average of the local particle charge profile.
Therefore, the sorting of the particles in the electric field also affects the local average charge.
The positively charged particles moved in the negative $y$ direction, and the negative ones in the positive $y$ direction.

Figures~\ref{fig:q_slice} and~\ref{fig:q_surf} reveal a bipolar charged particle flow.
The charge profile is nearly symmetric to $y/D=$~0.5.
Close to the left wall, the particles carry positive and, close to the right wall, negative charge, while the particles in the bulk are, on average, nearly neutral. 
The positive and negative charge increases from $z/D$~=~0.1 toward $z/D$~=~0.9. 
The positive peak of the charge profile in slice $z/D$~=~0.9 is higher in magnitude than the negative peak at the opposite wall of the same slice.
However, according to the red and blue areas in figure~\ref{fig:q_surf}, a larger flow region carries an average negative charge.

The measured bipolar charge emanated from same-material contacts with other particles, the feeder, and the duct's walls.
During feeding, the particles contacted the feeder's screw, which was already covered with other adhering particles.
Since also the duct is made of PMMA, the flow-charged particles underwent only same-material contacts.
Same-material contacts, lacking a net direction of charge transfer, resulted in the observed bipolar charging.

%faraday
Finally, the two horizontal lines in figure~\ref{fig:q_slice} compare the derived charge to measurements of the Faraday pail (cf.~figure~\ref{fig:poexpa}).
During each PTV measurement, we recorded the mass and total charge of the powder captured by the Faraday pail.
The gray line in figure~\ref{fig:q_slice} gives the average charge over the 18 measurements (9 slices, each with and without the electric field) reported in this paper.
The standard deviation of the 18 Faraday readings of 0.043\,fC is less than the thickness of the gray line.
The orange line gives the corresponding average computed from the spatially resolved charge, which means the average of the charge profiles weighted by the local particle concentration.
The shaded area indicates the uncertainty of the average resolved charge.

Besides the errors of the spatially resolved measurements discussed in section~\ref{sec:acc}, the duct connecting the measurement section to the Faraday pail contributes to both method's discrepancies.
Because of the size of the operating equipment, the Faraday pail starts only 6\,$D$ after the end of the measurement section.
While the particles traverse through the connecting PMMA duct, they accumulate additional charge that affects the Faraday pail measurement.
Additionally, in contrast to the Faraday pail that captures all particles, the spatially resolved measurements miss particles in the wall's proximity.

Within these limits, the averages of the spatially resolved and the Faraday pail measurements agree well.
The spatially resolved measurements result in an average charge of -2.21\,fC, and the Faraday pail measurements in 0.49\,fC.
Despite the detected charge peaks, both methods return an overall nearly neutral flow.

However, the spatially-resolved charge peak is 76 times higher than the average measured by a Faraday pail.
Thus, a Faraday pail dramatically underestimates the flow's charge. 

\section{Conclusions}

This paper presents a novel in situ technology for measuring the electrostatic charge of particles in a turbulent duct flow.
Unlike existing methods that provide the sum of all particles' charges or are invasive, the developed technology spatially resolves the time-averaged charge profile across the flow.
By combining Particle Tracking Velocimetry (PTV) and electrode plates generating an electric field, the local particle charge is derived by comparing particle velocities and accelerations with and without the electric field.
We discuss the limitations of the new technique and compare the average derived from the spatially-resolved charge profile to the average charge provided by a Faraday pail.
Both methods agree well and indicate an overall nearly neutral flow.

However, the spatially-resolved results revealed a bipolar particle flow pattern.
The resolved charge peak was 76 times higher than the average particle charge detected by the Faraday pail.
The presented method opens up a new way to detect so far hidden charge peaks in powder operations and, thus, to enhance industrial safety.
Further, it provides the data required to validate and improve current simulations of powder flow charging.

The accuracy analysis points toward the future improvement of the new technique by reducing the length of the electric field.

\section*{Declarations}

\noindent\textbf{Ethical Approval:}
No studies on humans or animals were part of this work.
 
\noindent\textbf{Competing interests:}
The authors declare that they have no competing interests, or other interests that might be perceived to influence the results and/or discussion reported in this paper.
 
\noindent\textbf{Authors' contributions:}
W.X. developed the measurement system, with the help of S.J., based on the idea of H.G.
W.X. conducted and post-processed all experiments.
T.M. contributed to designing the test rig.
H.G. wrote the first draft of the manuscript.
All authors analyzed and interpreted the data, revised the article, and approved the submission.

\noindent\textbf{Funding:}
W.X., S.J., and H.G. have received funding from the European Research Council~(ERC) under the European Union's Horizon 2020 research and innovation program~(grant agreement No.~947606 PowFEct).
 
\noindent\textbf{Availability of data and materials:}
Raw and derived data supporting the findings of this study are available from the corresponding author upon reasonable request.

\bibliography{bibliography}

\end{document}